\documentclass{article}
\usepackage{spconf,amsmath,graphicx}
\usepackage{booktabs}
\usepackage{hyperref}  
\usepackage{multirow}
\usepackage{color}
\usepackage[cmyk]{xcolor}
\usepackage{subfigure}

\title{WeKws: A production first small-footprint end-to-end \\ Keyword Spotting Toolkit}
\name{Jie Wang$^{1,4}$, Menglong Xu$^{3,4}$, Jingyong Hou$^{2,4}$, Binbin Zhang$^{3,4}$, Xiao-Lei Zhang$^1$, Lei Xie$^2$, Fuping Pan$^3$}
\address{
	$^1$School of Marine Science and Technology, Northwestern Polytechnical University, Xi'an, China\\
        $^2$Audio, Speech and Language Processing Group (ASLP@NPU), \\School of Computer Science, Northwestern Polytechnical University, Xi’an, China \\
	$^3$Horizon Robotics, Beijing, China\\
	$^4$WeNet Open Source Community \\
	}
\begin{document}
%
\maketitle
\begin{abstract}
Keyword spotting (KWS) enables speech-based user interaction and gradually becomes an indispensable component of smart devices.
Recently, end-to-end (E2E) methods have become the most popular approach for on-device KWS tasks.
However, there is still a gap between the research and deployment of E2E KWS methods.
In this paper, we introduce WeKws, a production-quality, easy-to-build, and convenient-to-be-applied E2E KWS toolkit.
WeKws contains the implementations of several state-of-the-art backbone networks, making it achieve highly competitive results on three publicly available datasets.
To make WeKws a pure E2E toolkit, we utilize a refined max-pooling loss to make the model learn the ending position of the keyword by itself, which significantly simplifies the training pipeline and makes WeKws very efficient to be applied in real-world scenarios.
The toolkit is publicly available at https://github.com/wenet-e2e/wekws.
\end{abstract}
\section{Introduction}
\label{sec:intro}
Keyword spotting (KWS) is the task of detecting predefined keywords from continuous audio streams.
Wake-up word (WuW) detection, as a special task of KWS, has become a typical and indispensable component in internet of things (IoT) devices, e.g. smart speakers and mobile phones, enabling users to have a fully hands-free voice interactive user experience.
A WuW detection system needs to process streaming audio and runs locally and persistently on IoT devices.
Therefore, the system should have a small memory footprint and computational cost, while maintaining a low latency (real-time response) and high detection accuracy.

Thanks to the emergence of some neural networks with long-term modeling capabilities, end-to-end (E2E) based KWS methods recently have drawn much attention for their simplicity of training/decoding~\cite{maxpooling,shan2018attention,coucke2019efficient,hou2020mining,zhang2020re}.
Under the E2E setting, KWS becomes a keyword/non-keyword binary classification task.
Posteriors corresponding to different keywords are predicted directly by a model.
Then a system can easily detect keywords, by comparing the posteriors with manually defined thresholds of the keywords.
E2E based methods yield significant better performances over conventional methods~\cite{silaghi2005spotting,chen2014small,tang2018deep,wang2020wake,wang2021wake}.

However, the present E2E KWS methods still have defects.
Firstly, to train the model, they usually need a force-alignment procedure to get the start-end position of keywords in utterances~\cite{maxpooling,coucke2019efficient,hou2020mining,zhang2020re}.
Secondly, some of them rely on the entire input audio sequence to make decisions~\cite{bai2019time,zeng2019effective,mekonnen2022end}, so are not applicable to streaming tasks.
In addition to the above shortcomings of the present E2E KWS methods, there is a lack of an open-source toolkit like Wenet~\cite{wenet} in the field of small-footprint KWS to bridge the gap between research and production. There are several good speech-processing toolkits, e.g. Kaldi~\cite{povey2011kaldi}, Fairseq~\cite{ott2019fairseq} and Honk~\cite{tang2017honk}, that have implemented KWS as part of their features. However, they are either too complex~\cite{povey2011kaldi, ott2019fairseq} or far away from real production in design~\cite{tang2017honk}.

In this paper, to tackle the above-mentioned problems, we introduce WeKws, a production-oriented lightweight E2E KWS toolkit. The key advantages of WeKws are as follows.

\noindent $\bullet$
\textbf{Alignment-free}: WeKws is an alignment-free E2E toolkit. There is no need to use an automatic speech recognition (ASR) or speech activity detection (SAD) system to get the alignments or ending timestamps of keywords, which significantly simplifies the KWS training pipeline.

\noindent $\bullet$
\textbf{Production ready}:
We try our best to bridge the gap between research and production when we design WeKws.
Wekws uses causal convolutions to achieve streaming KWS.
All modules of WeKws conform to the requirements of TorchScript\footnote{https://pytorch.org/docs/stable/jit.html}. 
Therefore, a model trained with WeKws can be exported by Torch Just In Time (JIT), transformed to Open Neural Network Exchange \footnote{https://github.com/onnx/onnx} (ONNX) format, and easily adopted in many deployment environments.

\noindent $\bullet$
\textbf{Light weight}:
WeKws is designed specifically for E2E KWS with clean and simple codes, and it only depends on PyTorch. The trained model is lightweight, and is able to run on embedded devices.

\noindent $\bullet$
\textbf{Competitive results}:
WeKws achieves competitive results on several public KWS benchmarks compared with other recently proposed KWS systems. 
Moreover, WeKws does not require alignments and decoding graphs, while the comparison systems are unable to do so, making their training pipelines complicated and rely on other heavy toolkits.

\section{WeKws}
\label{sec:wekws}

\subsection{System design}
\begin{figure}[t]
    \centering
    \includegraphics[width=0.42\textwidth]{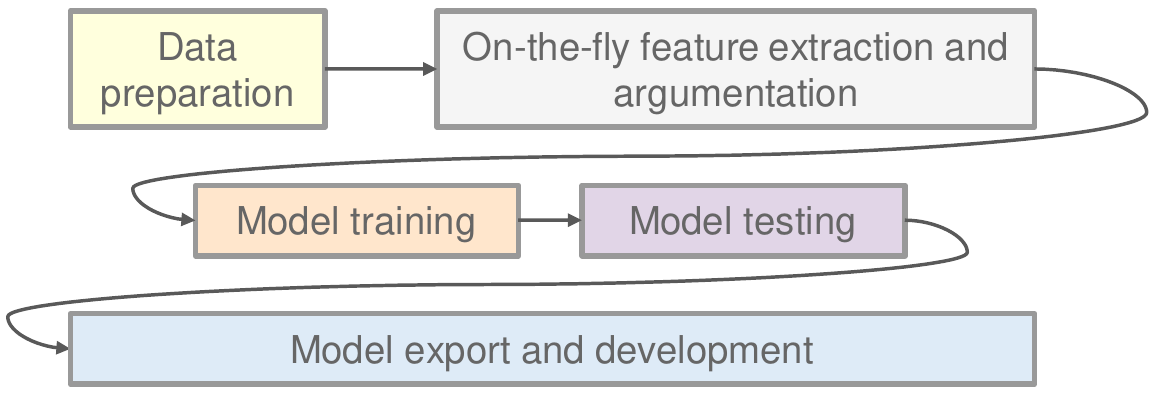}
    \caption{System design of the proposed WeKws toolkit.}
    \label{fig.system_design}
\end{figure}

As shown in Figure~\ref{fig.system_design}, our proposed WeKws is designed with three layers.

\subsubsection{Layer 1: Data preparation module and an on-the-fly feature extraction and argumentation.}
In the data preparation module, all we need to do is to prepare an audio list and utterance-level keyword category labels required for model training.
WeKws uses an on-the-fly feature extraction and argumentation procedure.
Each utterance is first re-sampled to a specific sampling rate, followed by speed perturbation and Mel-filter bank feature extraction. Feature-level Specaugment~\cite{Specaugment} are applied to generate the input of the KWS model.
It is worth noticing that the above procedure is performed dynamically and on-the-fly in each training mini-batch.
Compared with traditional offline feature preparation, this on-the-fly procedure not only helps save disk usage but also enriches the diversity of training examples across different epochs, which improves the robustness of the model.

\subsubsection{Layer 2: Model training and testing.}
We have provided an easy-to-use platform for researchers to develop and evaluate their own algorithms.
At present, we have implemented a unified model architecture with several popular KWS backbones and a refined max-pooling KWS training objective which greatly simplified the KWS training pipeline.
The detailed model architecture and the refined max-pooling training objective will be introduced in the next two sub-sections.
We also provided standard KWS testing criteria. Competitive results on several public KWS benchmarks were achieved with the proposed model and training method.

\subsubsection{Layer 3: Model exportation and development.}
The trained model supports TorchScript and ONNX exportation, and therefore can be easily applied to different platforms. Currently, we support deploying the WeKws production model on three mainstream platforms, i.e., x86 server, android, and Raspberry Pi. Furthermore, both \textit{float32} model and quantized \textit{int8} model are supported in WeKws, where the quantized \textit{int8} model can increase the inference speed when hosted on embedded devices such as the ARM-based Android and Raspberry Pi platforms. 

\subsection{Model architecture}
\begin{figure}[t]
    \centering
    \includegraphics[width=0.36\textwidth]{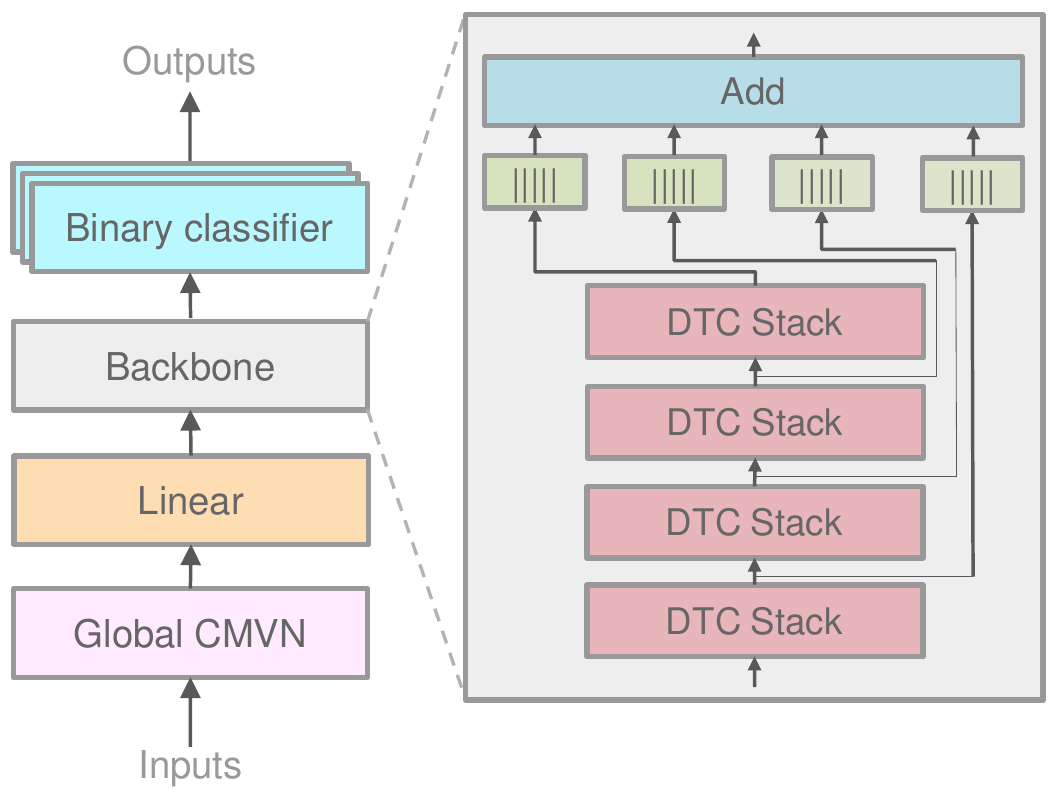}
    \caption{A WeKws implementation with MDTC as the backbone, where the depthwise temporal convolution (DTC) as a basic block. The details of the MDTC network can be found in~\cite{MDTC}.}
    \label{fig.model}
\end{figure}

The overall model architecture of WeKws is shown in Figure~\ref{fig.model}.
It consists of four parts, starting with a global cepstral mean and variance normalization (CMVN) layer to normalize the input acoustic features to a normal distribution. This is followed by a linear layer which maps the dimensions of the input features to the required dimensions.
Then there is a backbone network, which can be recurrent neural network (RNN), temporal convolutional network (TCN)~\cite{coucke2019efficient} or multi-scale depthwise temporal convolution (MDTC)~\cite{MDTC}.
At the end of the model, there are several binary classifiers, each of which
has a single output node with a sigmoid activation to predict the posterior probability of the corresponding keyword.
As~\cite{hou2020mining} did, for each keyword, we add an independent binary classifier after the backbone network to deal with the scenarios with multiple keywords. Note that the backbone network is shared by these binary classifiers.

WeKws current supports the following three backbones:
1) RNN or its improved version LSTM~\cite{maxpooling}, which has been widely used in speech recognition and other speech tasks;
2) TCN~\cite{coucke2019efficient} or its lightweight version depthwise separable TCN (DS-TCN) with dilated convolutions to increase the receptive field;
3) MDTC~\cite{MDTC}, an advanced backbone network that extracts multi-scale feature representation efficiently.
Note that as we aim to achieve a streaming KWS system, we use causal convolutions in all convolution-based neural networks.

\subsection{Refined max-pooling based training objective} \label{sec:max_pooling}
A refined max-pooling loss is used as the training objective of WeKws, which is formulated as:
\begin{equation}
\mathcal{L} = -\left[y_i \ln p^*_i + \left(1-y_i\right) \ln \left(1-p^*_i\right)\right]
\end{equation}
with
\begin{equation}
p^*_i = \max_j(\mathbf{p}_{ij}),\ \ \ \forall j=m+1, m+2, \cdots, N.
\end{equation}
where $\mathbf{p}_{ij}$ is the predicted posterior probabilities, $m$ is the minimum duration frames of the keyword, and $N$ is the number of frames of the $i$-th utterance in each training mini-batch.
Note that the minimum duration $m$ is calculated statistically on all positive samples in the training set.

By using the max-pooling loss, the model learns the ending timestamps of the keyword automatically, and thus gets rid of the dependence on alignments or the ending timestamps of the keyword.
Specifically, for the positive samples, the max-pooling loss will only optimize the frames with the highest posterior probabilities, and ignore the other frames.
For the negative samples, the max-pooling loss will minimize the frames with the highest posterior probabilities, therefore, the posterior probabilities of all frames in the negative samples will be minimized.

\section{Experiments}
\label{sec:experiment}
We conducted experiments on several public KWS benchmarks. We will introduce our experimental setup and report the results in this section.
\begin{table}[tp]
\centering
\caption{FRR(\%) comparison of the proposed WeKws and other KWS systems with FAH fixed at 0.5, on keywords ``Hi xiaowen" and ``Nihao wenwen".}
\label{table:delta}
\vspace{6pt}
\scalebox{0.95}{
\begin{tabular}{lccc}
\hline
System  & \#Params &   Hi xiaowen     & Nihao wenwen \\
Wang at al.~\cite{wang2021wake}   & $57 \mathrm{k}$         & ${0.7}$        & ${0.5}$ \\
WeKws                             & $153 \mathrm{k}$        & $\mathbf{0.50}$ & $\mathbf{0.43}$ \\
\hline
\end{tabular}}
\end{table}

 \begin{table}[tp]
\centering
\caption{FRR (\%) comparison of the proposed WeKws and other KWS systems on keyword ``Hey snips".}
\label{table:delta2}
\vspace{6pt}
\begin{tabular}{lccc}
\hline
System    & \#Params   & FAH=0.5   &  FAH=1 \\
Zhang at al.~\cite{zhang2020re}    & $244 \mathrm{k}$  & 3.53               & 2.82 \\
Coucke at al.~\cite{wavenet}       & $222 \mathrm{k}$  & $\mathbf{0.12}$    & ${-}$ \\
WeKws                              & $153 \mathrm{k}$  & $\mathbf{0.12}$    & $\mathbf{0.08}$ \\
\hline
\end{tabular}
\end{table}
\begin{table}[t]
 \centering
 \caption{Comparison of the proposed WeKws and other KWS systems on Google Speech Command.}
 \vspace{6pt}
 \label{tab:gsc}
 \begin{tabular}{ccc}
 \hline System                        &      \#Params        &     Accuracy (\%)       \\
 Zhang at al.~\cite{gsc3}             &      109K            &       97.20    \\
 Ding at al.~\cite{gsc4}               &      404K            &       97.56    \\
 WeKws                                &      158K            &       $\mathbf{97.97}$    \\
 \hline
 \end{tabular}
 \end{table}
\subsection{Experimental setup}
We used Mobvoi (SLR87)~\footnote{https://www.openslr.org/87}, Snips~\footnote{https://research.snips.ai/datasets/keyword-spotting}
and Google Speech Command (GSC)~\cite{googlespeechcommand}
corpora to evaluate our proposed WeKws.
The Mobvoi is a Mandarin corpus designed specifically for WuW tasks.
There are two keywords in the corpus, each having about $36$K utterances.
The number of non-keyword utterances is about $183$K. We followed the same setup as~\cite{wang2021wake} to use the corpus.
The Snips is a crowd-sourced WuW corpus.
The keyword of the corpus is ``Hey snips", and there are around $11$K keyword utterances and $86.5$K non-keyword utterances in the corpus. 
The detailed corpus information can be found in~\cite{coucke2019efficient} and we followed the recommended setup in~\cite{coucke2019efficient} to carry out our experiments.
The GSC corpus consists of $64, 721$ one-second-long recordings of 30 words by 1881 different speakers. Following the typical setup like~\cite{MDTC}, we used utterances in \textit{validation\_list.txt} and \textit{testing\_list.txt} as validation and testing data, respectively, and used the other utterances as training data.

We used 40-dimensional Mel-filter bank (Fbank) features, with 25~ms window size and 10~ms window shift, as the model input.
We used Adam, with an initial learning rate of $1e-3$ and an L2 weight decay of $1e-4$, as an optimizer for the model training. The batch size was set to $128$ utterances. The proposed WeKws was trained with 80 epochs.
We selected 30 best models on the development set, from the total 80 models saved after each epoch, and then averaged them to obtain the final model.

\subsection{Experimental results}
\subsubsection{WeKws \textit{vs.} other KWS systems}

In this section, we have evaluated our proposed WeKws on three public benchmarks and compared WeKws with other competitive KWS systems proposed recently~\cite{zhang2020re,wang2021wake,wavenet,gsc3,gsc4}.
We used MDTC with the best setup in~\cite{MDTC} as the backbone of our model, and the refined max-pooling loss introduced in Section~\ref{sec:max_pooling} to train the model.
In Table~\ref{table:delta}, compared with an LF-MMI-based KWS system which relies on a graph-based decoding algorithm~\cite{wang2021wake}, our WeKws improves the false rejection rate (FRR) relatively by around 28\% and 14\% on ``Hi xiaowen" and ``Nihao wenwen", respectively, on the Mobvoi corpus.
In Table~\ref{table:delta2}, we compare our WeKws with two E2E methods on the ``Snips" corpus. Our WeKws achieves competitive results with the two methods, and a smaller memory footprint than the latter. Moreover, the comparison methods need a SAD system to get the ending timestamps of keywords in training, which is complicated, while our method avoids to do so. In Table \ref{tab:gsc}, WeKws achieves the best result, compared with two recent KWS methods on the GSC dataset.
The above results support the superiority of our WeKws.

\subsubsection{Comparison of different training objectives}
To evaluate the effect of max-pooling loss, we compared it with several representative loss functions in literature.
The \textbf{vad-mean} loss mainly follows~\cite{coucke2019efficient, zhang2020re}, in which the authors utilize the SAD technique to get the ending timestamps of the keyword in the positive samples, and maximize the posteriors of all frames within a given time interval around the ending timestamp. In our implementation, we maximized the average of the posteriors within the given time interval, and the time interval was set to 5 frames.
The \textbf{vad-max} loss follows~\cite{maxpooling}, where a max-pooling function is used to pick the most informative frame within the alignment of the keyword. Different from~\cite{maxpooling}, we restrict the picked frame to be around the ending timestamp of the keyword, and the range was set to 40 frames.
The \textbf{weakly-constraint} loss was proposed by~\cite{hou2020mining}. It only uses the constraint of the ending timestamp at the early stages of training and relaxes the constraint after a preseted training epoch. In our experiments, we used the constraint in the first 5 epochs.

\begin{figure}[t]
\centering
\subfigure[Hi xiaowen]{
\begin{minipage}[b]{0.48\linewidth}
\centering
\includegraphics[width=1\textwidth]{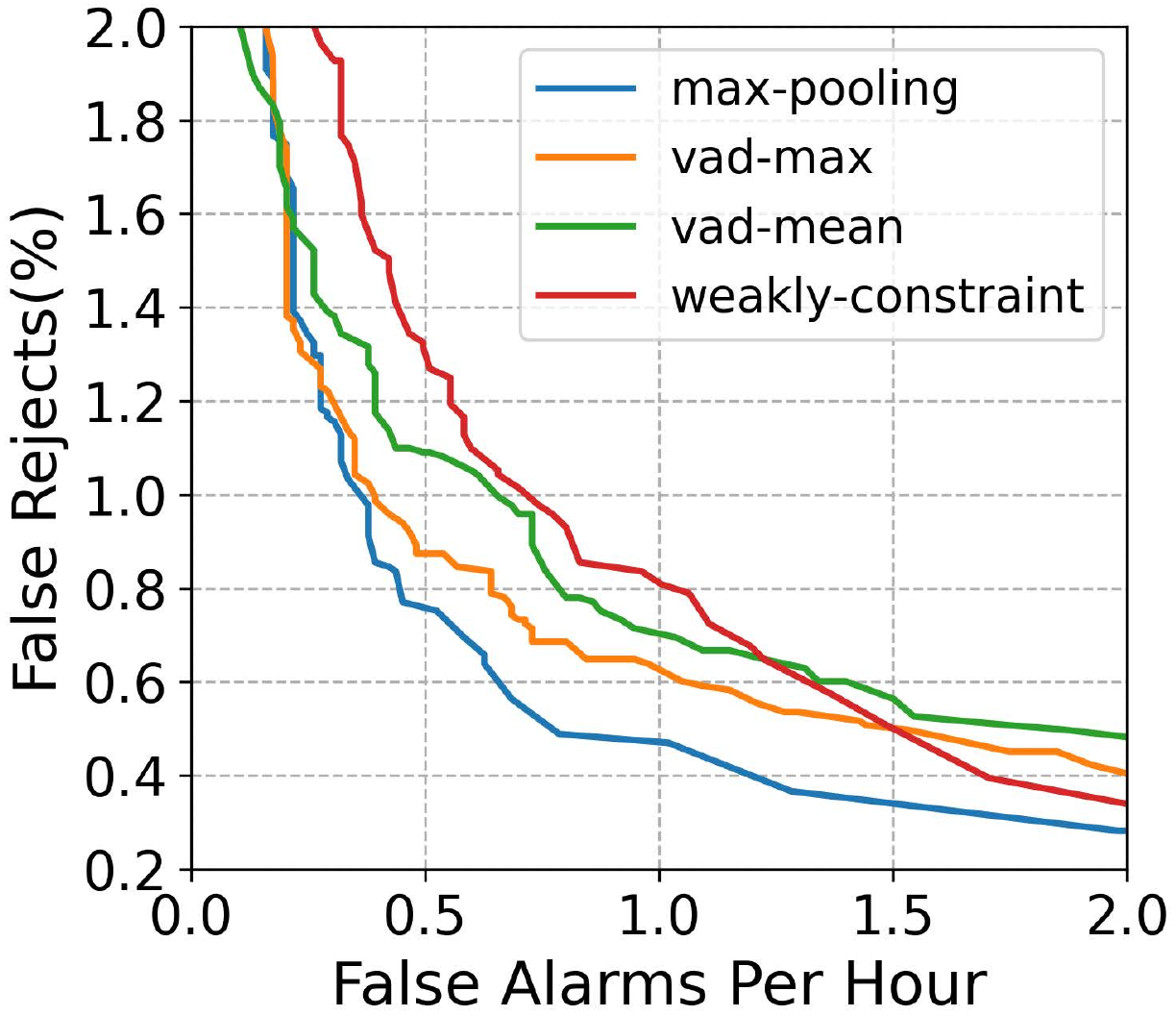}
\end{minipage}%
}
\subfigure[Nihao wenwen]{
\begin{minipage}[b]{0.48\linewidth}
\centering
\includegraphics[width=1\textwidth]{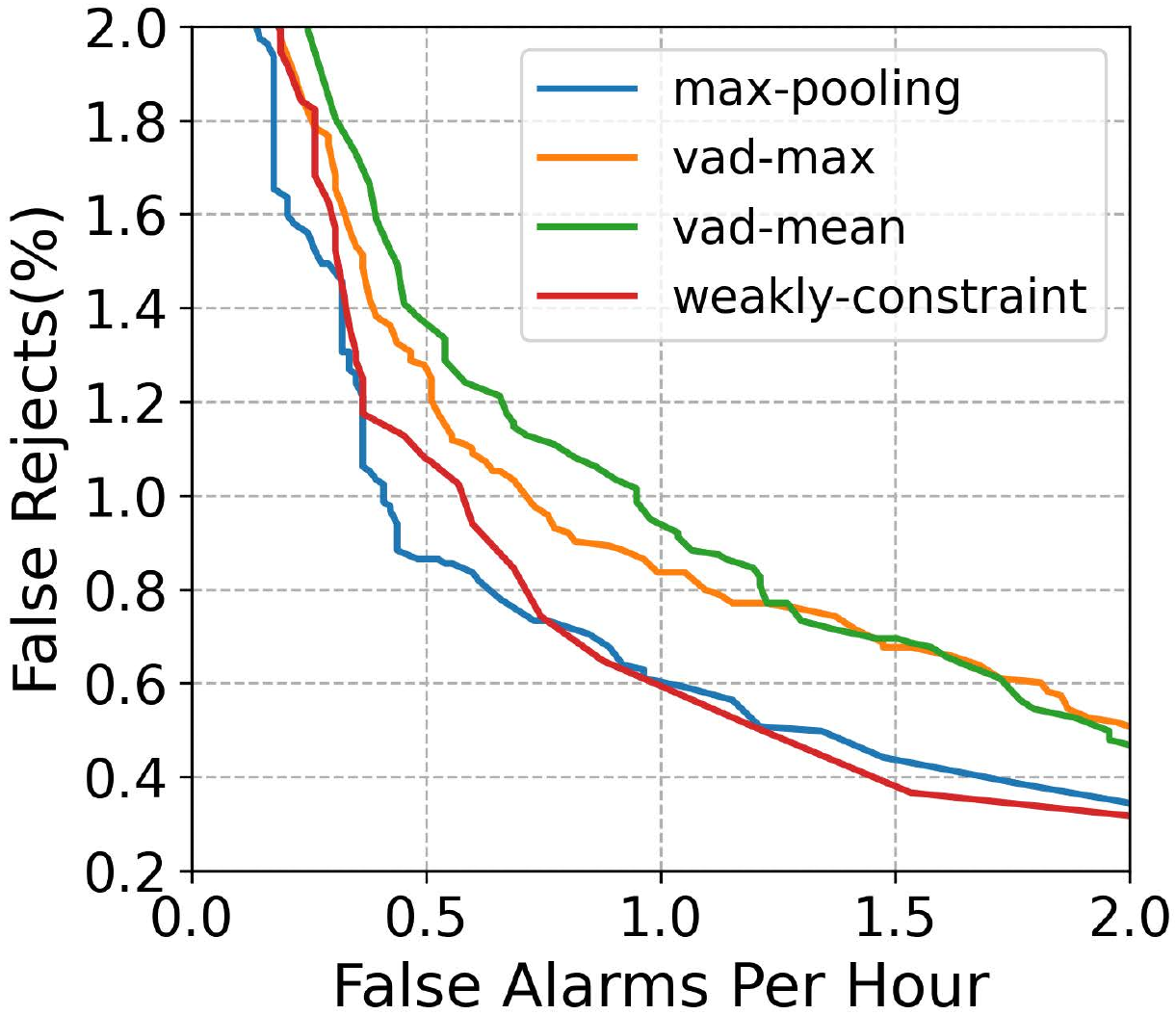}
\end{minipage}%
}
\centering
\caption{DET curve comparison of different models trained with max-pooling, vad-max, vad-mean, and weakly-constraint training objectives, on keywords ``Hi xiaowen" and ``Nihao wenwen".}
\label{fig.det}
\end{figure}

All experiments in this section were based on the DS-TCN backbone, and the comparison results are shown in Figure~\ref{fig.det} and Table~\ref{tab:loss}.
From Figure~\ref{fig.det}, we see that the \textbf{max-pooling} loss achieves the best performance in most cases on the two keywords.
More specifically, from Table~\ref{tab:loss} we see that the \textbf{max-pooling} loss overcomes all comparison losses at 0.5/1 False Alarm per hour. 
In addition, it is worth mentioning that the \textbf{max-pooling} loss does not use any additional supervised information, which makes WeKws a simple and effective KWS system.
 \begin{table}[t]
 \centering
 \caption{FRR(\%) comparison of different training objectives with FAH fixed at 0.5/1.0, on keywords ``Hi xiaowen" and ``Nihao wenwen".}
 \vspace{6pt}
 \label{tab:loss}
 \begin{tabular}{c|cc}
 \hline Training objective            &    Hi xiaowen    &     Nihao wenwen  \\
 \hline vad-max       &    $0.87$/$0.64$   &     $1.28$/$0.84$     \\
 \hline weakly-constraint &    $1.32$/$0.81$   &     $1.08$/$0.65$     \\
 \hline vad-mean         &    $1.09$/$0.70$   &     $1.38$/$0.95$     \\
 \hline max-pooling      &    $\mathbf{0.75}$/$\mathbf{0.47}$   &     $\mathbf{0.87}$/$\mathbf{0.61}$     \\
 \hline
 \end{tabular}
 \end{table}
 \begin{table}[t]
 \centering
 \caption{FRR(\%) comparison of different backbones with FAH fixed at 0.5/1.0.}
 \vspace{6pt}
 \label{tab:model}
 \begin{tabular}{c|c|cc}
 \hline Backbone   &   \#Params  & Hi xiaowen       & Nihao wenwen        \\

 \hline  \hline TCN     &    2M        &    $0.63/0.42$   &     $0.69/0.44$     \\
 \hline DSTCN   &    287K      &    $0.75$/$0.47$   &     $0.87$/$0.61$      \\ 
 \hline GDSTCN  &    124K      &    $0.84$/$0.36$   &     $0.64$/$0.35$      \\ 
 \hline MDTC    &    153K      &    $\mathbf{0.50}$/$\mathbf{0.22}$   &     $\mathbf{0.43}$/$\mathbf{0.29}$     \\
 \hline
 \end{tabular}
 \end{table}
\subsubsection{Comparison of different backbone networks}
To further verify the effectiveness of the \textbf{max-pooling} loss, we applied it to multiple different backbone networks such as dilated TCN~\cite{TCN}, grouped DS-TCN~\cite{gdsconv}, and MDTC~\cite{MDTC}.
The comparison results are listed in Table~\ref{tab:model}. It shows that all backbone networks obtain good results with the \textbf{max-pooling} loss. From Table~\ref{tab:model}, we also see that the depthwise separable convolution significantly reduces the number of parameters with a small or no performance degradation.
In addition, we see that MDTC achieves the best trade-off between performance and model footprint.
Compared with DSTCN, MDTC uses almost half of the parameters to achieve $33.3\%$, $53.2\%$, $50.6\%$, and $52.5\%$ relative reduction on FRR at $0.5/1.0$ false alarm per hour on ``Hi xiaowen" and ``Nihao wenwen" respectively.


\section{Conclusions}
\label{sec:conclusions}
In this paper, we present a production-quality, easy-to-build, and convenient-to-be-applied open-source E2E KWS toolkit called WeKws.
With a pure E2E training objective and elaborate designed modules, WeKws can be trained, exported and applied on different platforms easily.
WeKws helps to bridge the gap between research and production of KWS methods and provide an easy-to-use platform for researcher and engineer.
Evaluations on three KWS benchmarks demonstrate that WeKws achieves competitive or better results compared with many state-of-the-art KWS methods.

\ninept
\bibliographystyle{IEEEbib}
\bibliography{refs}
\end{document}